# Leveraging Social Media Data and Artificial Intelligence for Improving Earthquake Response Efforts

Kalin Kopanov [0009-0001-8725-2643], Velizar Varbanov [0009-0007-9762-7924],

Tatiana Atanasova [0000-0002-6084-3179]

Institute of Information and Communication Technologies – Bulgarian Academy of Science, Sofia, Bulgaria
```
kalin.kopanov@iict.bas.bg
velizar.varbanov@iict.bas.bg
tatiana.atanasova@iict.bas.bg
```

**Abstract.** The integration of social media and artificial intelligence (AI) into disaster management, particularly for earthquake response, represents a profound evolution in emergency management practices. In the digital age, real-time information sharing has reached unprecedented levels, with social media platforms emerging as crucial communication channels during crises. This shift has transformed traditional, centralized emergency services into more decentralized, participatory models of disaster situational awareness. Our study includes an experimental analysis of 8,900 social media interactions, including 2,920 posts and 5,980 replies on X (formerly Twitter), following a magnitude 5.1 earthquake in Oklahoma on February 2, 2024. The analysis covers data from the immediate aftermath and extends over the following seven days, illustrating the critical role of digital platforms in modern disaster response. The results demonstrate that social media platforms can be effectively used as real-time situational awareness tools, delivering critical information to society and authorities during emergencies.

## 1 Introduction

The transformative role of social media in modern society is profoundly evident in its capacity to disseminate information swiftly during emergencies. Platforms such as X (formerly Twitter), Facebook, Telegram and TikTok transcend their traditional roles as channels for social interaction to become critical tools in emergency management. These platforms facilitate the rapid spread of crucial information, vital for initiating timely response actions. When accurately collected and analyzed, data from these sources significantly enhance situational awareness, improve decision-making processes, and expand geographic and contextual understanding in crisis situations.

The integration of Artificial Intelligence (AI) with social media has revolutionized disaster management strategies further. AI enhances the processing and accuracy of data derived from social media, enabling quicker and more precise responses. This integration between AI and social media platforms facilitates more efficient situational assessments, improves resource allocation, and strengthens disaster response coordination. AI-driven tools are particularly adept at filtering out non-essential information, prioritizing critical data, and ensuring the rapid dissemination of verified updates - key for managing public safety and responses in disaster-stricken areas.

The increasing dependence on social media for creating immediate situational awareness tools for natural hazards, security incidents, and public health crises highlights its growing importance. Moreover, there is an escalating demand for real-time disaster damage assessments that extend beyond the post-disaster phase. This paper explores how social media platforms can be effectively utilized as real-time situational awareness tools to provide critical information to the public, government organizations, and non-governmental organizations (NGOs) during ongoing disasters.

## 2      Related work

Currently, disasters trigger substantial spikes in social media activity, as individuals share updates about their situations and experiences, and seek information to make informed decisions. This increase is not exclusive to platforms like X/Twitter or limited to the developed world [1].

Karimiziarani's comprehensive review [2] underscores the pivotal role of social media analytics in earthquake disaster management, highlighting its capabilities in enhancing response and recovery efforts. Social media data has proven instrumental in facilitating rapid damage assessment, pinpointing disruptions in critical infrastructure, and bolstering coordination during emergency responses. The integration of this data with seismic monitoring systems has emerged as a best practice, enabling more effective and timely interventions. Additionally, the review details the use of sentiment analysis for monitoring public anxiety levels during seismic events, which assists in addressing community concerns proactively. Machine learning algorithms are also noted for their efficacy in real-time detection of events, significantly improving the speed and accuracy of responses to seismic activities.

Academic research has shown the potential of social media for the real-time detection and monitoring of natural disasters. User-generated content on these platforms can act as early indicators, greatly aiding in rapid response and resource allocation. Particularly, X/Twitter's real-time capabilities allow for the immediate detection of events such as earthquakes. Users' posts (tweets) provide critical data that enable the deployment of algorithms to monitor and identify disaster occurrences. For instance, an advanced classifier analyzes post (tweet) features like keywords, word count, and context to detect specific events. Furthermore, a probabilistic spatiotemporal model is applied to determine the event's epicenter and path, utilizing techniques such as Kalman filtering and particle filtering for precise location estimation. This approach has proven exceptionally effective in Japan, where an X/Twitter-based earthquake report-ing system achieves a high detection rate of 96% for significant seismic events, providing notifications faster than traditional broadcast methods by the Japan Meteorological Agency (JMA). This system exemplifies how social media can transform disaster response mechanisms, enhancing both speed and accuracy [3].

The systematic review by Acikara, Xia, Yigitcanlar, and Hon on the role of social media in disaster response [4] highlights a comprehensive analysis spanning from January 2012, reflecting on a decade of significant digital advancements. This study, sourced from a detailed search on Google Scholar, began with an array of keywords such as "disaster risk reduction", "disaster response", "emergency response", and "social media". The intent was to thoroughly capture the scope of social media's integration into disaster risk management and response. The initial yield of 867 articles was refined to 421 peer-reviewed, English-language journal articles. Further examination for alignment with the study's goals reduced the pool to 183 articles. A meticulous full-text review, which prioritized articles utilizing actual social media data, distilled the selection to 102 relevant pieces. These were categorized based on their strategies for addressing the critical challenges of disaster response, showcasing social media's pivotal role in fostering situational awareness and collective intelligence amidst crisis.

## 3  Challenges and opportunities

The exploitation of social media analytics in earthquake response operations embodies a significant interplay between considerable challenges and promising opportunities. This convergence of digital communication platforms represents a fundamental transformation in how disaster management can harness real-time, user-generated content to effectively mitigate the impacts of catastrophic events. Furthermore, it will highlight opportunities for future implementation of location awareness and situational awareness to enhance support and response strategies. In the following subsections, we cover five principal challenges and opportunities; however, this is not an exhaustive list, and further challenges and opportunities undoubtedly exist and merit consideration.

### 3.1  Communication Infrastructure

The resilience of communication infrastructures is critically tested during earthquakes when traditional networks are often compromised. Depending on the earthquake's magnitude and severity, communication can be partially affected, temporarily disabled, or suffer minimal damage. These impacts are usually most severe near the earthquake's epicenter, although the wider region may experience the earthquake without significant damage to its communication infrastructure. In such scenarios, affected individuals may promptly begin interacting and posting updates about their status on social media. This immediate exchange of information underscores the inherent vulnerability of traditional communication networks and necessitates the de-velopment of robust, alternative communication systems that can operate independently of traditional frameworks.

### 3.2 Data Management Capabilities

The overwhelming influx of data during disasters poses significant challenges in its collection, processing, and analysis. Efficiently managing this data requires sophisticated computational approaches that can quickly filter through vast volumes of information to identify and prioritize actionable insights. This scenario presents an excellent opportunity to harness the full spectrum of AI technologies, including Natural Language Processing (NLP) techniques and Large Language Models (LLMs). These AI tools are well-equipped to handle the noisy, unstructured nature of social media content, which often includes multilingual content, sarcasm, and context-specific nuances, adding layers of complexity to data processing.

Furthermore, the multilingual aspect of social media data presents a considerable challenge. While NLP can achieve remarkable success in processing English, it often underperforms in languages with less available training data or those characterized by linguistic complexities, such as Arabic.

### 3.3 Geotagging

The use of geolocation data in social media analytics can significantly enhance the specificity and effectiveness of disaster response efforts. However, it is important to note that when users upload images and videos to social media platforms, metadata, including location data, is often stripped away by default. While users have the option to add geotags manually, these tags may not always be accurate, potentially leading to misdirection or false filtering by location data. Additionally, by default, most applications have the geolocation option turned off, which further limits the availability of precise location data. Despite these challenges, when available and accurate, geotag data can be incredibly valuable, providing snapshot information about who is responding and what is happening in the areas affected by an earthquake. This can be crucial for coordinating timely and effective disaster response efforts.

### 3.4 Application Programming Interfaces (APIs)

Access to social media data typically involves the use of APIs provided by the platforms. These APIs allow for the retrieval of data streams based on specific queries. However, significant challenges arise due to API constraints, including "pull" rather than "push" data access models, rate limits, and restrictive Terms of Service (ToS). Furthermore, social media companies often impose technological barriers to prevent data exploitation for commercial purposes or competitive development, complicating the acquisition of timely, relevant and voluminous data.

To proactively address these limitations, it is crucial to stay informed about updates to the ToS and rate limits of various social media platforms. Monitoring these changes can help in adapting strategies and tools to comply with legal requirements while maximizing data access. Additionally, developers should anticipate potential API constraints by creating advanced solutions ahead of time, which can be quickly deployed when needed. For instance, in early 2023, X (formerly Twitter) announced significant changes to its API access policy, which included the introduction of new pricing tiers and the elimination of the previously free API access. These revisions underscore the importance of staying agile and responsive to external platform changes that could affect data access and application functionality. Where legally permissible, developers might also consider designing AI-powered tools for data mining that can navigate or even leverage these API limitations to enhance data acquisition capabilities effectively. Such preemptive measures and innovations can significantly mitigate the impact of API restrictions on social media data utilization.

### 3.5 Seamless Integration with Emergency Management Systems

Integrating social media analytics tools into existing emergency management infrastructures presents a unique set of challenges, particularly in ensuring compatibility and enhancing the functionality of legacy systems. Some countries, like the United States, have developed sophisticated emergency management systems and utilize various platforms to enhance public communication and response strategies. For instance, the United States Geological Survey (USGS, https://earthquake.usgs.gov/) and the European Mediterranean Seismological Centre (EMSC, https://emsc-csem.org/) provide up-to-date information on seismic activities through dedicated sections on their websites and social media channels. These organizations manage robust web pages and social media outlets to disseminate timely and accurate information.

## 4 Experimental part

Our methodology involves an analysis of user-generated content and interactions on X (formerly Twitter), immediately following the Oklahoma earthquake which occurred on February 2, 2024, at 23:24:28 local time [5], and extending over the subsequent seven days (until February 9, 2024). This research utilized two types of queries: one based on keywords, using "Oklahoma" and "earthquake", and another based on location, covering a 250 km radius from the epicenter where the earthquake's impact was felt within Oklahoma State.

The keyword-based query yielded a total of 7,987 results, comprising 2,610 posts and 5,377 comments. The location-based query returned 1,988 results, including 652 posts and 1,336 comments. After merging these results and removing duplicates, a total of 1,075 entries were omitted. The final dataset thus contained 8,900 results, consisting of 2,920 posts and 5,980 comments. It is important to note that while our dataset likely includes most, if not all, posts from the analyzed period, the comments collected represent only the most recent or relevant ones - approximately 27% of the initially noted 22,111 comments as of February 10, 2024. Furthermore, it is worth noting that our statistics (for posts and comments) show the dataset recorded 325,221 reposts, almost 2 million likes, and 223 million views. These figures represent the initial data before any processing and filtering were applied.

As expected, we observed the highest user activities within the first 24 hours, accounting for approximately 85% of all posts and replies (comments). To better visualize the temporal spread of activity, we employed a Locally Weighted Scatterplot Smoothing (LOWESS)[1] approach for charting this data (fig. 1). By fitting simple models to localized subsets of the data, LOWESS captures the stochastic nature of the underlying processes without assuming a specific global model form. This method is particularly advantageous for our analysis as it allows the depiction of both the central tendency and dispersion of the data over time, thus accommodating the inherent variability and providing a more nuanced view of user engagement trends. The smoothed visualization underscores significant patterns while mitigating the distraction of short-term fluctuations, thereby facilitating a more intuitive interpretation of the temporal dynamics at play.

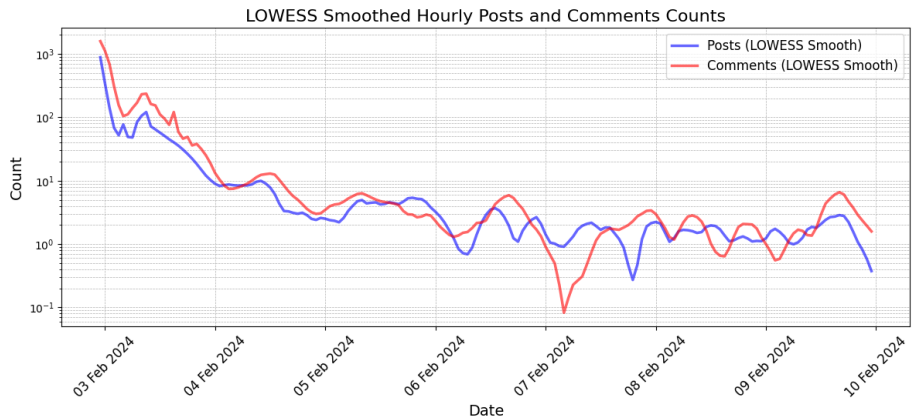

**Fig. 1.** Temporal spread of user activity following the Oklahoma Earthquake from February 2 to February 9, 2024

By tracking user activity and leveraging AI-driven natural language processing (NLP), we aim to filter out irrelevant information, thereby enhancing the focus on critical data that can complement existing sources such as applications, news, and seismological data. For our research, we chose to process the data with XLM-RoBERTa, which, according to our previous tests [6], demonstrated the best overall performance in recognizing location entities.

The analysis yielded significant insights, with 8,402 locations being identified from 4,914 posts and replies (comments), representing approximately 55% of the total 8,900 entries analyzed. Notably, similar to prior findings, approximately 83% of these locations were identified within the first 24 hours following the earthquake. Further refinement of the data revealed that 5,663 entries, or 67.4% of all detected locations, were directly associated with Oklahoma State and the related earthquake events.

---

[1] A non-parametric regression method that combines multiple regression models in a k-nearest-neighbor-based meta-model.

Subsequent steps involved cleaning the data to remove exact duplicates, which reduced the number of location mentions to 324. A more nuanced grouping process was then applied, where similar location references (e.g. "Tulsa County", "Tulsa OK", "Tulsa Oklahoma" and "West Tulsa") were consolidated into a single representative term ("Tulsa"). This consolidation resulted in the identification of 144 unique locations within Oklahoma State, as illustrated in Figure 2. This streamlined representation allows for a more focused analysis of the impact areas and facilitates more effective communication of the results.

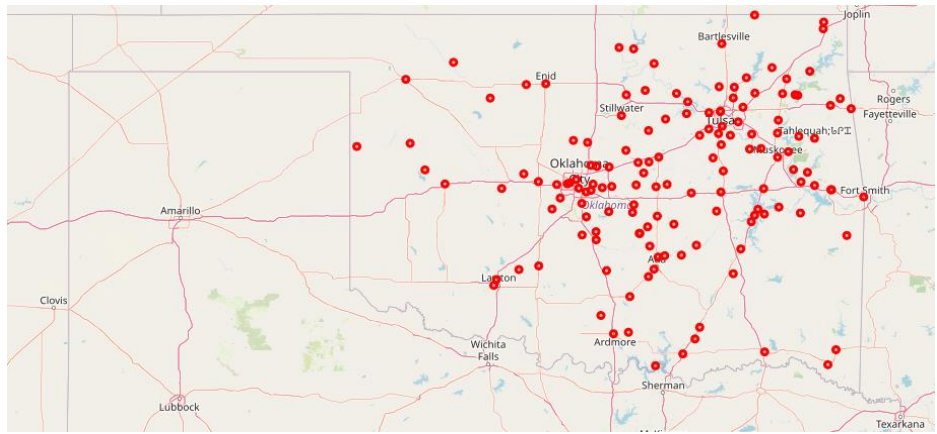

**Fig. 2.** Visualization of the 144 unique locations within Oklahoma State

In a separate test, we utilized AI, specifically OpenAI's GPT-4, and a LLM, to identify the coordinates of locations within Oklahoma State. This approach enabled GPT-4 (gpt-4-0125-preview) to accurately recognize 137 locations, achieving a 95% accuracy rate. However, it missed 7 locations where the exact location was unclear or not commonly recognized, such as Eufaula Lake or Eufaula Dam. This application of AI and LLMs technologies underscores their potential in enhancing the precision of geospatial analyses in our research.

Additionally, we enhance our monitoring capabilities by employing keyword analysis to detect text that signals events requiring attention or immediate response. For example, phrases such as "felt it", appearing 585 times, "shook", mentioned 257 times, and "damage", noted 144 times, suggest potential areas of concern that may warrant further investigation. Similarly, less frequent but critical terms like "explosion" (8 mentions), "crash" (7), and "fire" (6) indicate more urgent situations requiring prompt attention.

## 5      Conclusion

The integration of social media and AI into earthquake disaster management, as demonstrated in this study, could significantly enhances situational awareness and response efficiency. The rapid dissemination of user-generated content immediately following the Oklahoma earthquake highlights the importance of social media as a real-time data source. Utilizing advanced NLP techniques for geolocation identification, we were able to accurately filter and analyze critical data. This could not only improves decision-making and resource allocation but also strengthens coordination among emergency responders and aid organizations.

Despite challenges like data accuracy and information overload, the benefits of leveraging social media for disaster response are evident. The volume of data generated within the first 24 hours is particularly valuable, offering a crucial window for rapid response that can substantially mitigate the impact of disasters. Moving forward, enhancing the integration of these technologies in disaster management strategies is essential for developing more responsive and resilient systems, capable of protecting and informing communities during critical times.

## 6      Acknowledgments

This work was supported by the National Science Program "Security and Defense", which has received funding from the Ministry of Education and Science of the Republic of Bulgaria under the grant agreement no. Д01-74/19.05.2022.